\documentclass[aps,reprint,article,prx]{revtex4-1}
\usepackage{amsmath}
\usepackage{dcolumn}
\usepackage{graphicx}
\usepackage{bm}
\usepackage{float}
\usepackage{color}
\usepackage{ulem}

\begin{document}
	
	\title{Comparative studies on unconventional superconductivity in Cr$_3$Ru compounds with bcc and A15 structures}
	\author{Zhengyan Zhu, Ying-Jie Zhang, Yiwen Li, Qing Li}\thanks{liqing1118@nju.edu.cn}\author{Wen Duan}\author{ Hai-Hu Wen}\thanks{hhwen@nju.edu.cn}
	\affiliation{Center for Superconducting Physics and Materials, National Laboratory of Solid State Microstructures and Department of Physics,  Collaborative Innovation Center for Advanced Microstructures, Nanjing University, Nanjing 210093, China}
	\date{\today}
	
	\begin{abstract}
	Chromium (Cr) is a transition metal element with 3$d$ orbital electrons. In most compounds containing Cr, due to the correlation effect, twofold features of localization and itinerancy are expected. The localization gives rise to a magnetic moment, while the latter exhibits as the effective coherent weight for conductivity. Here we report the physical properties of Cr$_3$Ru compounds with bcc or A15 structures by using multiple experimental tools. The resistivity measurements show sharp superconducting transitions at $T_{\rm c}$ = 2.77 K and $T_{\rm c}$ = 3.37 K for the bcc and A15 structures, respectively. A high residual resistivity exists for both samples with the mean-free-path in the scale of about 2 nm. Magnetization measurements also show rather narrow transitions, with a clear hump structure at high temperatures ($T_{\rm c}$ $\le$ $T$ $\le$ 300 K), which may be ascribed to the remaining antiferromagnetic spin fluctuations. A pronounced second peak effect has been observed in magnetization hysteresis loops in the superconducting state only for samples with bcc structure. The specific heat coefficient reveals a clear jump at $T_{\rm c}$. We find that s-wave gaps can be adopted to fit the low temperature specific heat data of both samples yielding ratios of about $2\Delta/k_{\scriptscriptstyle B}T_{\rm c}$ $\approx$ 3.6, indicating a moderate pairing strength. Interestingly, the Wilson ratios $R_{\scriptscriptstyle W} = A\chi_0 / \gamma_n $ are 3.81 and 3.62 for the bcc and A15 phases, suggesting a moderate correlation effect of conducting electron in the normal state. Besides, for samples with A15 structure, another specific heat anomaly occurs at about 0.85 K and is sensitive to magnetic fields. By applying high pressures, both system will exhibit an enhancement of $T_{\rm c}$ with a rate of about 0.019 K/GPa and 0.013 K/GPa for the bcc and A15 phases, respectively. We also conduct tunneling spectrum measurements on these samples, and found that the coherence peaks are strongly smeared out with a gap in the scale of about 0.3 $\sim$ 0.5 meV. The strong suppression to the coherence peaks may be ascribed to the strong scattering seen from the resistivity measurements. Our combinatory results point to an unconventional superconductivity in these Cr based compounds.
	\end{abstract}
    \pacs{--------} \maketitle
	
	\section{Introduction}
	Cuprates \cite{LaBaCuO} and iron-based superconductors \cite{LaOFFeAs}, as unconventional superconductors with high critical temperatures ($T_{\rm c}$), all have strongly correlated 3$d$ electrons. The relationship between unconventional superconductivity and strong correlation effect of 3$d$ electrons with magnetic interactions and spin fluctuations has aroused great interest in the past decades. In recent years, lots of research have been conducted on unconventional superconductivity in other materials which contain 3$d$ transition-metal elements \cite{NaCoO2,nickelate,NiSTM,bulkNi,AV3Sb5,CsV3Sb5,MnP}. Chromium is a kind of 3$d$ transition-metal elements, and the ionic state of Cr has a strong magnetic moment. Thus Cr-based superconducting compounds may have rich and novel electronic behavior. For compounds containing 3$d$ transition-metal chromium, many works have been carried out based on a class of CrAs-based superconductors. CrAs itself is an antiferromagnet with $T_{\rm \scriptscriptstyle N}$ at about 270 $\pm$ 10 K \cite{CrAs}. When applying a pressure of about 0.8 GPa, CrAs exhibits superconductivity with a $T_{\rm c}$ of about 2 K \cite{CrAspressure}. Then a big family of CrAs-based superconductors with a quasi-one-dimensional (Q1D) crystal structure was discovered. There are two kinds of materials, one kind is $A_2$Cr$_3$As$_3$ ($A$ = alkali metal), including Na$_2$Cr$_3$As$_3$ ($T_{\rm c}$ $\approx$ 8.6 K) \cite{Na2Cr3As3}, K$_2$Cr$_3$As$_3$ ($T_{\rm c}$ $\approx$ 6.1 K) \cite{K2Cr3As3}, Rb$_2$Cr$_3$As$_3$ ($T_{\rm c}$ $\approx$ 4.8 K) \cite{Rb2Cr3As3}, and Cs$_2$Cr$_3$As$_3$ ($T_{\rm c}$ $\approx$ 2.2 K) \cite{Cs2Cr3As3}, the other is $A$Cr$_3$As$_3$, including KCr$_3$As$_3$ ($T_{\rm c}$ $\approx$ 5 K) \cite{KCr3As3} and RbCr$_3$As$_3$ ($T_{\rm c}$ $\approx$ 7.3 K) \cite{RbCr3As3}. There are accumulated evidence to show that some of these materials may be candidates for multiband superconductivity with spin-triplet electron pairing state \cite{CrAstriplet,RbCr3As3multiband,K2Cr3As3triplet}. In addition, the study of superconductivity in Cr-based materials always attracts much attention. Pr$_3$Cr$_{10-x}$N$_{11}$ was found as an example of possible unconventional superconductors with strong electron correlations \cite{Pr3Cr10-xN11}. The ternary compound Cr$_2$Re$_3$B with $\beta$-Mn-Type structure was also reported to be a superconductor with $T_{\rm c}$ = 4.8 K \cite{Cr2Re3B}. And a new discovery of superconductivity with $T_{\rm c}$ $\approx$ 7 K at external pressure indicates the quantum criticality and unconventional superconductivity in CrB$_2$ \cite{CrB2}. Actually in the early years, some binary Cr-based compounds were reported as superconductors \cite{CrSC,transitionmetalSC,SCferromagnetic,SC3d,CrRe,CrRu,pressureA15}, but detailed studies on physical properties are lacking. Moreover, 4$d$ transition element Ruthenium (Ru) is one of iron's relatives with strong ferromagnetism and itinerant electron magnetism \cite{SrRuO3,RRu2Si2}. In addition to Sr$_2$RuO$_4$ \cite{Sr2RuO4} as a candidate for spin-triplet superconductors \cite{Sr2RuO4triplet}, other Ru-based compounds also have unusual superconductivity, such as Ru$_7$B$_3$ as a non-centrosymmetric superconductor \cite{Ru7B3} and LaRu$_2$P$_2$ with an intricate Fermi surface topology \cite{LaRu2P2}. All of these inspire us to conduct studies of physical properties in chromium-based Cr-Ru binary alloys.	
	
	According to previous report, with the substitution of Ru, the antiferromagnetic (AF) order of Cr vanishes and superconductivity emerges gradually. And 25$\%$Ru is in the region where the two phases coexist \cite{CrRu}. Another study on different Cr$_{1-x}$Ru$_x$ alloys has discussed the relationship between the AF spin density wave (SDW) and superconductivity \cite{Cr1-xRux}. They found that the two phases are exclusive. And there may be spin fluctuations near the boundary of the two phases. The band structure calculations of Cr$_3$Ru reveal multiple sets of Fermi surfaces which are mainly contributed by the 3$d$ orbital electrons of chromium. Interestingly, there are two kinds of structures for Cr-Ru alloys. One is the body-centered cubic (bcc) structure where Cr atoms and Ru atoms arbitrarily occupy each site \cite{CrRu}. And the other is the cubic A-15 ($\beta$-W) crystal structure achieved after long time annealing the bcc Cr$_3$Ru \cite{A15}. And the A15 structure was reported in favor of the existence of superconductivity in many compounds \cite{A15SC}. According to the literature \cite{pressureA15}, the slight excess of Ru is more conducive to obtain pure phase of Cr$_3$Ru with A15 structure, thus we synthesized two kinds of polycrystals with a ratio of Cr : Ru = 72 : 28 and carried out comparative tests with various experimental tools.
	
	In this paper, we report the comparative studies of Cr$_3$Ru superconductors with bcc and A15 structures. The crystal structure, chemical composition, magnetization and resistance measurements are carried out to confirm the high quality of the samples with $T_{\rm c}$(bcc) = 2.82 K and $T_{\rm c}$(A15) = 3.39 K, respectively. The resistance measurements both show strong residual resistance with negligible magnetoresistance. And the magnetization measurements under high magnetic field represent clear hump structures at around 150 K for two kinds of samples. The sharp specific heat jumps for both samples can be fit well with s-wave gap of $2\Delta/k_{\scriptscriptstyle B}T_{\rm c}$ around 3.6. But the Wilson ratios are very large, about $R_{\scriptscriptstyle W}$(bcc) = 3.81 and $R_{\scriptscriptstyle W}$(A15) = 3.62 in these two samples. In magnetization hysteresis loops (MHLs), a second peak effect has been observed only in samples with bcc structure. Also, with the increase of the external pressure, the $T_{\rm c}$ enhances linearly for two samples. In the tunneling spectra, due to strong scattering, the coherence peaks are smeared out with a gap around 0.3 $\sim$ 0.5 meV.

	\section{Experiments}	
	The polycrystalline samples of Cr$_3$Ru with bcc structure were prepared by using arc-melting method. The chromium (99.98$\%$, Alfa Aesar) and ruthenium (99.9$\%$, Alfa Aesar) with a mole ratio of 72 : 28 were weighed, ground and pressed into tablet in a glove box filled with argon. Then the tablet was melted in the arc-melting furnace filled with pure argon. To improve the homogeneity of samples, each ingot was remelted at least three times. After that, the Cr-Ru alloy with bcc structure was obtained. Then the ingots of Cr$_3$Ru was placed into an evacuated quartz tube and annealed at 1000$^{\circ}$C for one month. The samples with A15 structure can then be obtained.	
	
	The X-ray-diffraction (XRD) measurements were carried out on a Bruker D8 Advanced diffractometer with the Cu-K$\alpha$ radiation. The TOPAS 4.2 software was used to refine the crystal structures by Rietveld analysis. The micrographs and the chemical composition analysis of the samples were taken by a Phenom ProX scanning electron microscope (SEM). The magnetization was measured using a SQUID-VSM (Quantum Design). The resistivity was measured by the standard four-probe method using a physical property measurement system (PPMS 16T, Quantum Design). And the resistance down to 0.4 K was measured with an additional $^3$He insert. The specific heat was measured with the thermal-relaxation method by another option of PPMS with the $^3$He insert. The resistivity data under high pressure were collected by our PPMS equipment adding a diamond-anvil-cell (cryoDAC-PPMS, Almax easyLab) module with a four-probe van der Pauw method. The scanning tunneling spectra (STS) measurements were carried out in a scanning tunneling microscope (USM-1300, Unisoku Co., Ltd.).

	\section{Results}
	
	\subsection{Sample characterization}
	\begin{figure}
		\includegraphics[width=8cm]{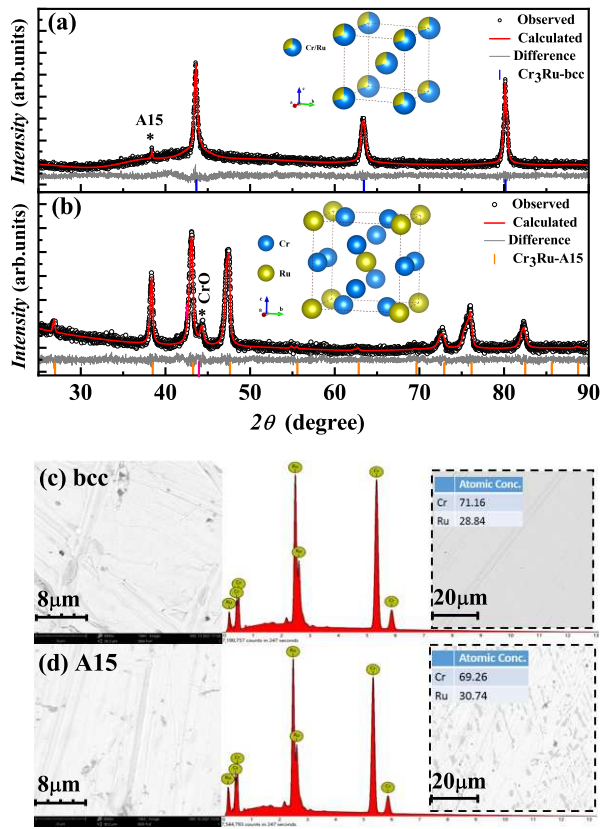}
		\caption{XRD patterns (black circles) with Rietveld refinement fitting curves (red lines) of (a) Cr$_3$Ru with bcc structure and (b) Cr$_3$Ru with A15 structure. The insets show the schematic crystal structures respectively. SEM images and the corresponding energy dispersive spectra (EDS) of (c) Cr$_3$Ru with bcc structure and (d) Cr$_3$Ru with A15 structure.}\label{fig1}
	\end{figure}	
	Figure~\ref{fig1}(a) and~\ref{fig1}(b) show the XRD patterns for two kinds of polycrystalline samples of Cr$_3$Ru and the corresponding Rietveld fitting curves. The Rietveld refinement indicates that the samples are assigned to be the cubic structure (space group: $Im\bar{3}m$ for bcc and $Pm\bar{3}n$ for A15) with the lattice parameters of 2.937 $\AA$ and 4.637 $\AA$, respectively. The schematic crystal structures are also plotted in the insets in which the blue and yellow spheres represent Cr and Ru, respectively. In the inset of Fig.~\ref{fig1}(a), Cr and Ru occupy the positions as bcc structure in a random way, but after high temperature annealing, the crystal structure becomes A15. In Fig.~\ref{fig1}(a), except for a minor peak of Cr$_3$Ru with A15 structure at around $2\theta = 38^{\circ}$, all other peaks belong to bcc structure. In Fig.~\ref{fig1}(b), except for a minor peak of impurities of Cr-O, all other peaks belong to Cr$_3$Ru with A15 structure. The scanning electron microscope (SEM) images with backscattered electron (BSE) mode of the two samples are shown on the left of Fig.~\ref{fig1}(c) and~\ref{fig1}(d). They respectively illustrate the morphology of the sample surfaces within a 30 $\mu$m area of view. We can only see some scratches caused by physical cuts on the surfaces of the samples. In order to determine the exact element composition of the samples, we also perform energy dispersive X-ray spectra (EDS) on a larger scale. The element compositions for two samples are Cr : Ru = 71 : 29 and 69 : 31 for bcc and A15 structure, respectively. The atomic ratios are close to our stoichiometric ratio.	
	\begin{figure}
		\includegraphics[width=7cm]{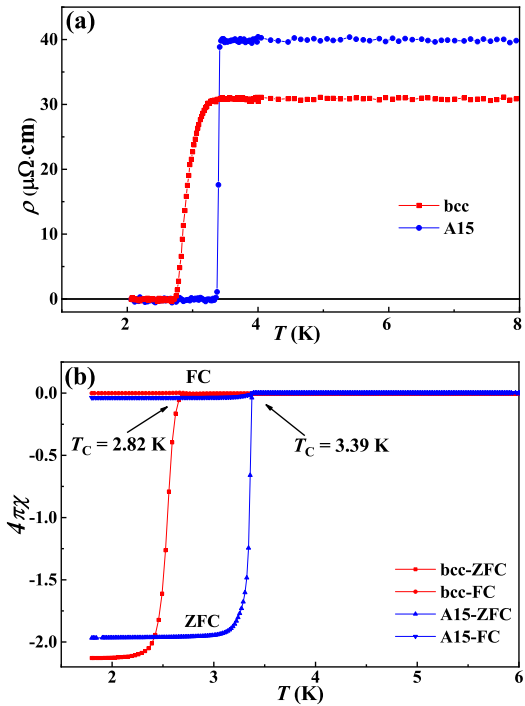}
		\caption{(a) Temperature dependence of resistivity for two kinds of samples. (b) Temperature dependence of magnetic susceptibility measured in both ZFC and FC modes for two kinds of samples with the applied magnetic field of 10 Oe.} \label{fig2}
	\end{figure}	

	The temperature dependence of resistivity at zero magnetic field for two kinds of samples are shown in Fig.~\ref{fig2}(a). The onsets of the superconducting transition temperatures are almost the same. Then with the decrease of temperature, the sample with A15 structure shows a very sharp superconducting transition. But the superconducting transition width is relatively wider for the sample with bcc structure. The obtained $T_{\rm c}$ from 10$\%$ of the normal state resistivity are 2.77 K for bcc, and 3.37 K for A15 structures, respectively. It should be pointed out that, according to previous reports \cite{CrRu}, the $T_{\rm c}$ of Cr$_3$Ru with bcc structure is significantly lower than that of A15 structure. The similar onsets of superconducting transition temperatures observed in our study may be the cause of the extra components with A15 structure in sample with bcc structure during the synthesis process. Fig.~\ref{fig2}(b) represents the temperature dependence of zero-field-cooled(ZFC) and field-cooled(FC) magnetization curves with the applied magnetic field of 10 Oe. Obtained $T_{\rm c}$ from magnetization are 2.82 K (bcc) and 3.39 K (A15). The superconducting volumes calculated from the magnetization data are both larger than 100$\%$ due to the demagnetization effect. All of these confirm the high quality of our samples.

	\subsection{Magnetic and Electrical transport properties}	
	\begin{figure}
		\includegraphics[width=7cm]{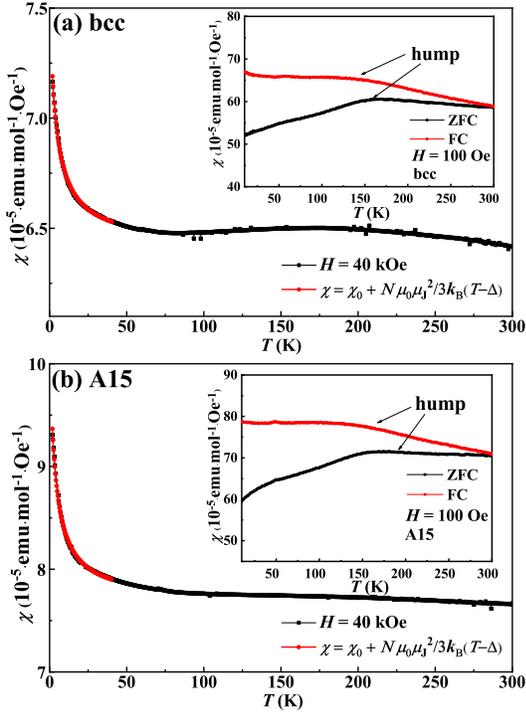}
		\caption{Temperature dependence of magnetic susceptibility from 1.8 K to 300 K under a magnetic field of 40 kOe for samples with (a) bcc and (b) A15 structures. The insets of (a) and (b) show the temperature dependencies of magnetic susceptibility measured with ZFC and FC modes at 100 Oe.} \label{fig3}
	\end{figure}	
	To investigate the magnetization behavior of Cr$_3$Ru under high magnetic field, we present the temperature dependent magnetic susceptibility curves under high fields in Fig.~\ref{fig3}(a) and~\ref{fig3}(b). We can see that the $\chi$-$T$ curves of both samples exhibit paramagnetic Curie-Weiss behavior in the low temperature region. And in high temperature range, the magnetic susceptibility does not change significantly with temperature. Note that the background signal near 50 K due to frozen oxygen has been subtracted by doing a Gaussian fitting. Then we fit them from 2 K to 50 K using the Curie-Weiss law, as shown by the red lines. For Cr$_{0.72}$Ru$_{0.28}$, the magnetic moments per unit molecule are 0.02 $\mu_{\scriptscriptstyle B}$/f.u. (bcc) and 0.027 $\mu_{\scriptscriptstyle B}$/f.u. (A15) for the two samples. We all know that Cr ions have a relatively large magnetic moment, but the magnetic moment of Cr-Ru alloy is very small. And the values of the magnetic susceptibility at zero temperature were estimated to be 3.56$\times$10$^{-4}$ emu/(mol$\cdot$f.u.) (bcc) and 3.82$\times$10$^{-4}$ emu/(mol$\cdot$f.u.) (A15), respectively. The insets in the figures represent the magnetic susceptibility with an applied magnetic field of 100 Oe. Both samples show broad hump structures in wide temperature regions centered around 150 K in the ZFC and FC curves. The hump structure centered around 150 K in the insets are similar to previous reports \cite{CrRu}. The humps may suggest the existence of AF ordering or spin fluctuations \cite{Cr1-xRux}. Since such humps are much weaker in the magnetization curve at higher magnetic fields, we believe that the humps in the magnetization curve may be related to the AF fluctuations.	
	\begin{figure*}
	    \includegraphics[width=15cm]{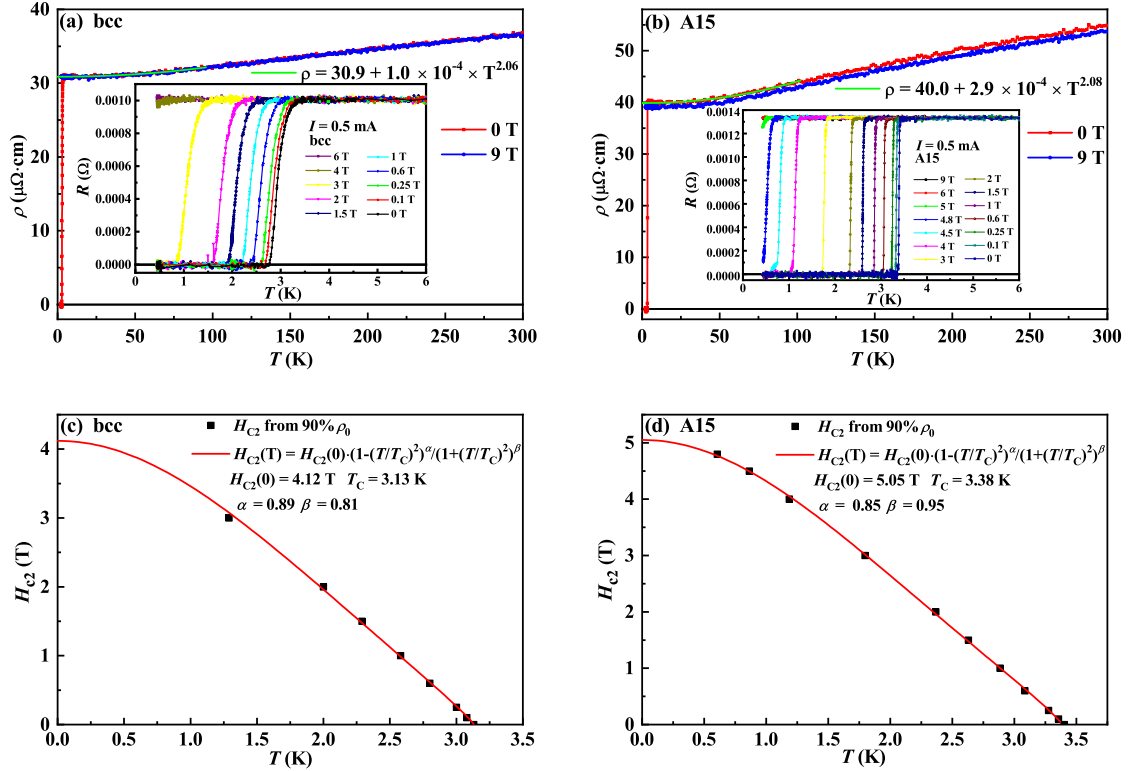}
	    \caption{Temperature dependence of resistivity from 2 K to 300 K under magnetic fields of 0 T and 9 T (a) for samples with bcc structure and (b) for samples with A15 structure. The insets of (a) and (b) show the temperature dependence of resistance under different magnetic fields from 0.4 K to 6 K. The upper critical field as a function of temperature (c) for samples with bcc structure and (d) for samples with A15 structure. The red lines show the fitting results obtained from the Ginzburg-Landau (GL) theory.} \label{fig4}
	\end{figure*}
	
	Figure~\ref{fig4}(a) and~\ref{fig4}(b) respectively represent the temperature dependence of resistivity for the two kinds of samples in the temperature range from 2 K to 300 K. Both samples exhibit sharp superconducting transitions at zero field. When the magnetic field up to 9 T is applied, the superconductivity is completely suppressed, and the resistivity shows normal metal behavior. It is found that $\rho$ exhibits a $T^2$ dependence below 100 K. The residual resistivity ratio ($\rho_{\scriptscriptstyle 300K}$/$\rho_{\scriptscriptstyle 2K}$, RRR) is 1.186 for the sample with bcc structure and is 1.377 for the sample with A15 structure. To our surprise, for both kinds of samples, the temperature dependence of resistance is very weak, and large residual resistances are present. In order to explore the possible causes of the residual resistivity, we conducted the following analysis of our data. First, the normal state resistivity at 0 T and 9 T overlap well in Fig.~\ref{fig4}(a) and~\ref{fig4}(b). This indicates that the magnetoresistance of the samples is almost negligible. We rule out the possibility of magnetic scattering. Then we measured the Hall resistance of the samples. Due to the large thickness of the samples and the high carrier concentration, the measured Hall resistance has large noise signals, and the carrier concentration can just be roughly estimated, that is in the order 10$^{23}$ /cm$^3$ \cite{Cr1-xRux}. We calculate the mean free path of the samples according to the formula $ l_{ab} = \hbar(3\pi^2)^{\frac{1}{3}}/e^2\rho_{\scriptscriptstyle 0}n^{\frac{2}{3}} $, where $\rho_{\scriptscriptstyle 0}$ is the residual resistivity at zero temperature and $n$ is the carrier concentration \cite{meanfreepath}. The mean free paths of both samples are very small, about 1 $\sim$ 2 nm. Therefore, the cause of the large residual resistance should be impurity scattering. As we can see from the SEM images of the two samples in the left side of Fig.~\ref{fig1}(c) and~\ref{fig1}(d), the alloy samples are very dense and have no discernible grains. Since no obvious grains are observed in the SEM images of the samples, the possibility of grain boundary scattering can be ruled out for the large residual resistivity. Considering all facts mentioned above, we argue that it may be due to some intrinsic scattering in samples. For example, the extra Ru atoms make the original crystal structure non-centrosymmetric. The insets of Fig.~\ref{fig4}(a) and~\ref{fig4}(b) show the temperature dependent resistance from 6 K down to 0.4 K under different magnetic fields. With the increase of applied magnetic field, the superconducting transition temperature decreases gradually. Fig.~\ref{fig4}(c) and~\ref{fig4}(d) represent the upper critical field at different temperatures. We obtain the values of upper critical field  from the $R$-$T$ curves by using criterions of 90$\%$ $R_{\rm n}$. By fitting the data of upper critical field $H_{\rm c \scriptscriptstyle 2}(T)$ with the Ginzburg-Landau theory, it is obtained that the upper critical field at zero temperature is $\mu_{\scriptscriptstyle 0}H_{\rm c \scriptscriptstyle 2}(0)$ = 4.12 T for samples with bcc structure and $\mu_{\scriptscriptstyle 0}H_{\rm c \scriptscriptstyle 2}(0)$ = 5.05 T for samples with A15 structure. Using the formula $\mu_{\scriptscriptstyle 0}H_{\rm c \scriptscriptstyle 2}(0) = \Phi_{\scriptscriptstyle 0}/2\pi\xi^2(0)$, the coherence lengths for the two kinds of samples are 9 nm (bcc) and 8 nm (A15), respectively. Considering the mean free path of 1 $\sim$ 2 nm, it is obvious that these superconductors are in the dirty. We all know that impurity scattering will affect the electron-phonon coupling and also be detrimental to the superconducting phase coherence. It is very unusual that these Cr-based compounds still exhibit superconductivity under such strong impurity scattering.
	
	\subsection{Specific heat}	
	\begin{figure*}
		\includegraphics[width=15cm]{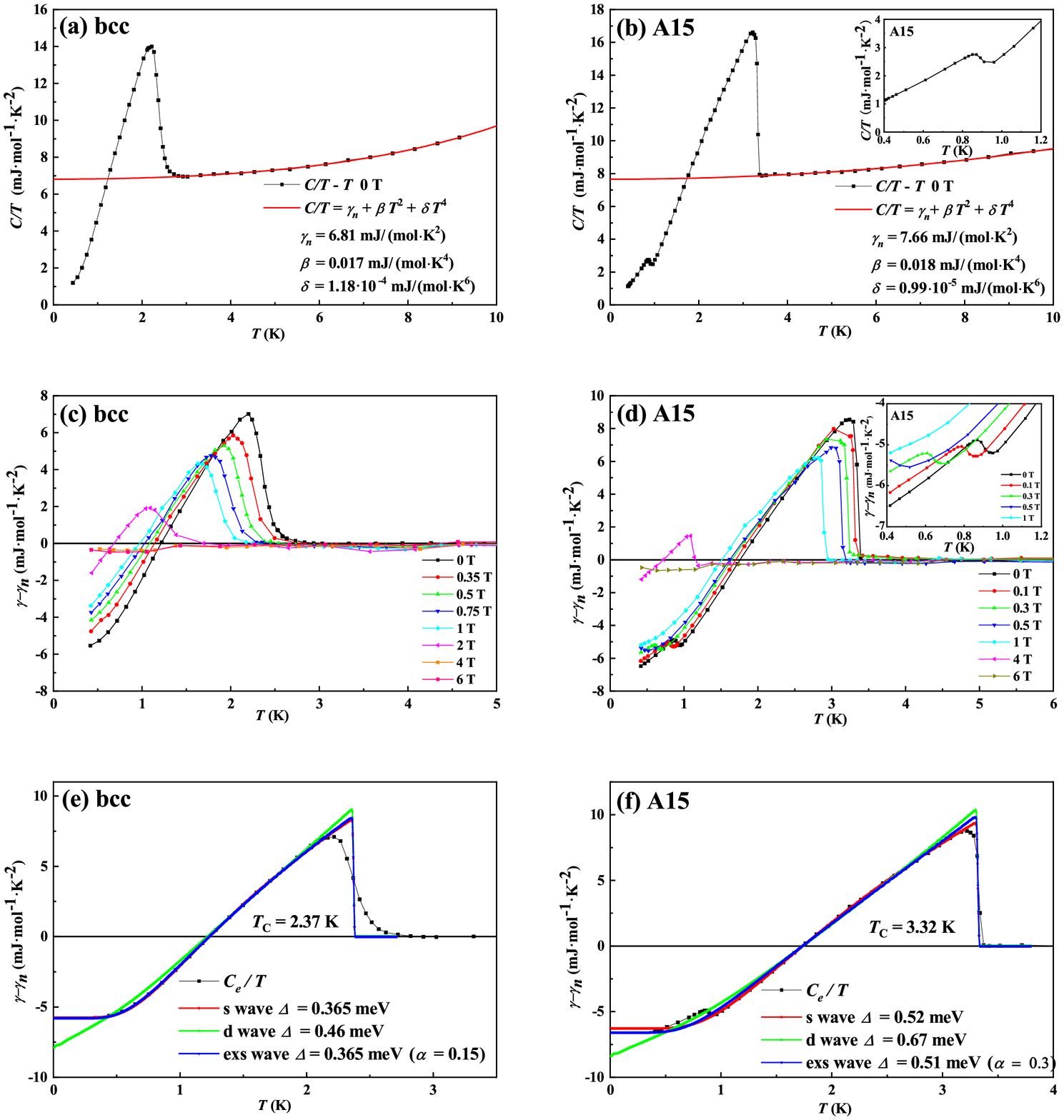}
		\caption{Temperature dependence of specific heat at zero magnetic field from 0.4 K to 10 K (a) for samples with bcc structure and (b) for samples with A15 structure. The red lines are the fitting curves of the normal state based on the Debye model. The inset of (b) shows the enlarged view of the anomaly around 0.85 K. The electronic specific heat under different fields (c) for samples with bcc structure and (d) for samples with A15 structure. The inset of (d) shows the temperature dependence of the anomaly under different magnetic fields. The temperature dependence of electronic specific heat coefficient at zero field and the fitting curves of different gap structures (e) for samples with bcc structure and (f) for samples with A15 structure.} \label{fig5}
	\end{figure*}	
	The temperature dependencies of specific heat of Cr$_3$Ru samples with bcc and A15 structures down to 0.4 K are given in Fig.~\ref{fig5}(a) and~\ref{fig5}(b). The data are plotted as $C/T$ versus $T$. Sharp specific heat jump can be seen at about 2.37 K for samples with bcc structure and 3.32 K for samples with A15 structure. For Cr$_3$Ru with bcc structure, the superconducting transition temperature determined by specific heat is lower than that determined by magnetization and resistivity. This is consistent with the behavior reported previously \cite{CrRu}. Surprisingly, for samples with A15 structure, another anomaly occurs at about 0.85 K, as shown clearly in the inset of Fig.~\ref{fig5}(b). We fit the data of the normal state using the equation $C_n/T = \gamma_n\,+\,\beta T^2\,+\,\eta T^4$, where $\gamma_n$ is the normal state electronic specific heat coefficient, called the Sommerfeld coefficient, and $\beta T^2+\eta T^4$, are the phonon contributions according to the Debye model. The normal state fittings all satisfy the law of entropy conservation. For Cr$_{0.72}$Ru$_{0.28}$, the Sommerfeld coefficient $\gamma_n$(bcc) = 6.81 mJ/(mol$\cdot$K$^2$) and $\gamma_n$(A15) = 7.66 mJ/(mol$\cdot$K$^2$) for the two kinds of samples. Using the equation $ \Theta_D = (12\pi^4k_{\scriptscriptstyle B}N_{\rm \scriptscriptstyle A}Z/5\beta)^{1/3} $, the Debye temperatures estimated here are about 485 K (bcc) and 476 K (A15) respectively, where $k_{\scriptscriptstyle B}$ is Boltzmann constant, $N_{\rm \scriptscriptstyle A}$ is Avogadro constant, $Z$ is the number of atoms in one unit cell. For Cr$_{0.72}$Ru$_{0.28}$, use $Z$ = 1. The specific heat jumps at $T_{\rm c}$, namely $\Delta C/\gamma_nT_{\rm c}$ with $\Delta C$ estimated by entropy conservation near $T_{\rm c}$, are about 1.2 for both of them. These are smaller than 1.43 which is predicted by the BCS theory in the weak-coupling limit. This finding is consistent with the weak phonon scattering predicted by the weak temperature dependent resistivity. The density of the states near the Fermi level obtained by the equation $N(\varepsilon_{\scriptscriptstyle F}) = 3\gamma_n/\pi^2k_{\scriptscriptstyle B}^2$ is 2.9 states/eV for samples with bcc structure and is 3.3 states/eV for samples with A15 structure. In comparison with the density of states on the Fermi surface from the band structure calculations \cite{AFLOW}, 5.2 states/eV for Cr$_6$Ru$_2$ with A15 structure, we can calculate the carrier effective mass $m^*/m_e$ = 5.1. A slightly large effective mass means the strong electron-electron interaction. Using the zero-temperature magnetic susceptibility estimated above and the Sommerfeld coefficient here, the Wilson ratio can be calculated as $R_{\scriptscriptstyle W} = \frac{4\pi^2k_{\scriptscriptstyle B}^2}{3(g\mu_{\scriptscriptstyle B})^2}\frac{\chi_0}{\gamma_n}$ =  $7.28\times$10$^4$ $\frac{\chi_0}{\gamma_n}$. For the two kinds of samples, the Wilson ratios are 3.81 and 3.62, respectively. These values are much larger than $R_{\scriptscriptstyle W}$ = 1 for the case of free-electron approximation. Therefore, this result implies strong electron correlations in these materials.
		
	Shown in Fig.~\ref{fig5}(c) and~\ref{fig5}(d) are the electronic specific heat under different magnetic fields for the bcc and A15 samples obtained by subtracting the normal-state background at zero field derived above. With increasing magnetic field, the specific heat jumps become broader for samples with bcc structure, but the specific heat jumps remain steep for samples with A15 structure. And the field induced specific heat coefficient increases gradually with the increase of magnetic field. In addition, the inset of Fig.~\ref{fig5}(d) shows an enlarged view of the second anomaly at 0.85 K under different magnetic fields for the A15 sample. It is clear that this anomaly shifts to lower temperatures with increasing magnetic field, thus is also sensitive to magnetic fields. The XRD on the A15 sample shows that the polycrystalline sample is quite pure, and the only impurity of CrO is not superconductive. We are not clear at this moment about the possible cause of the second specific heat anomaly in the A15 samples. It looks more like a superconducting transition, but it remains unclear whether it is due to an intrinsic transition, or due to some tiny impurity superconducting phase at low temperatures.
	
	In order to obtain more information about the superconducting gap structure, we use the BCS formula to fit the electronic specific heat in the superconducting state. The superconducting electronic specific heat is obtained by subtracting the normal state data following the Debye model. The superconducting electronic specific heat satisfies the entropy conservation. Based on the formula of specific heat given by the BCS theory, we use different kinds of gap structures to fit our data: a single s-wave gap $\Delta(T,\theta) = \Delta_0(T) $, a single d-wave gap $\Delta(T,\theta) = \Delta_0(T)\cos 2 \theta $, and a single extended s-wave gap $\Delta(T,\theta) = \Delta_0(T)(1+\alpha\cos 2 \theta) $ where $\alpha$ is the parameter that represents anisotropy. The optimized fitting parameters according to different gap structures are shown in Fig.~\ref{fig5}(e) and~\ref{fig5}(f) as different fitting curves. One can see that both the single isotropic s wave model and the single extended s-wave model can well describe the experimental data for two kinds of samples, even considering the anomaly around 0.85 K. These results indicate that one gap model is sufficient to describe the data. And the anisotropy of the superconducting gaps for the two samples is also very weak. The superconducting gaps are about 0.365 meV (bcc) and 0.52 meV (A15) for the two samples, respectively. The calculated ratios of $2\Delta/k_{\scriptscriptstyle B}T_{\rm c}$ are about 3.6 for the two samples, indicating a moderate pairing strength. We use the electronic specific heat integral to obtain the entropy, and then use the entropy integral to obtain the condensation energy of the superconductor. For Cr$_{0.72}$Ru$_{0.28}$, the condensation energy is 7.4 mJ/mol for samples with bcc structure, and 17.3 mJ/mol for samples with A15 structure.
	
	\subsection{Magnetization hysteresis loops and vortex properties}
	\begin{figure*}
		\includegraphics[width=14cm]{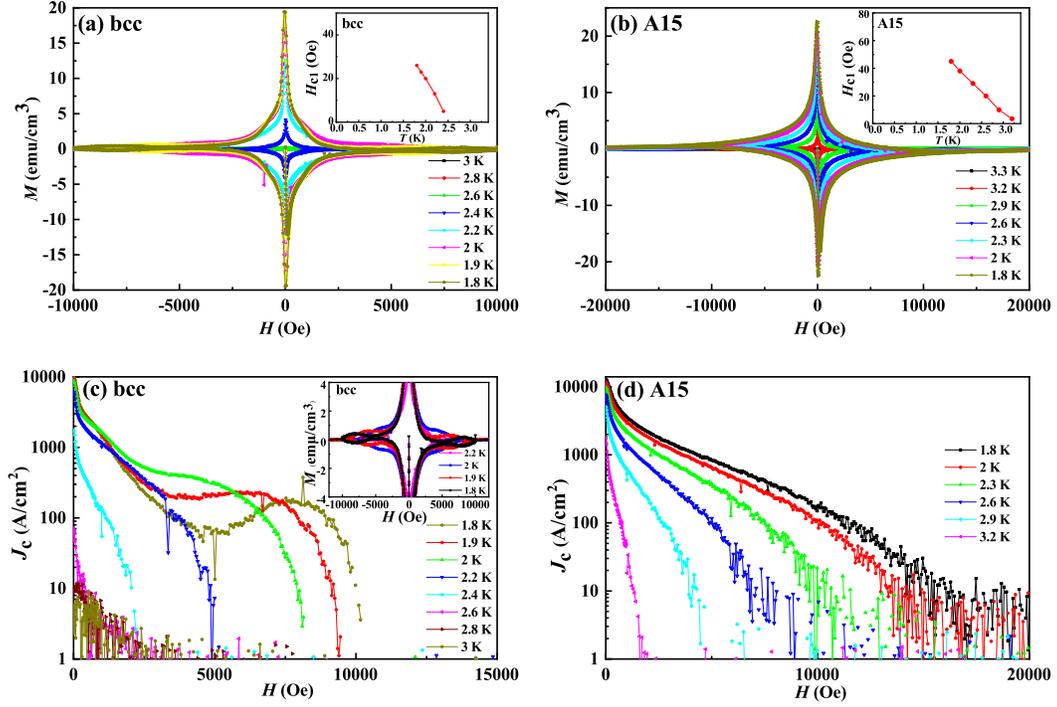}
		\caption{Magnetization hysteresis loops (MHLs) at different temperatures (a) for samples with bcc structure and (b) for samples with A15 structure. The insets of (a) and (b) show the temperature dependent lower critical fields. Magnetic field dependence of critical current density estimated by the Bean critical state model (c) for samples with bcc structure and (d) for samples with A15 structure. The inset of (c) shows the zoom-in view of MHLs with clear second peak effect.} \label{fig6}
	\end{figure*}	
	
    Figure~\ref{fig6}(a) and~\ref{fig6}(b) show the MHLs measured at different temperatures for the two kinds of samples. The insets show the lower critical field for the bcc and A15 samples, respectively. The lower critical fields at different temperatures are determined where $M$(H) deviates from the linear line corresponding to the Meissner state. The lower critical fields of the samples are very small, which means that the quantized flux can very easily enter the superconductors. The samples with bcc structure at different temperatures have a pronounced second magnetization peak effect as can be seen in the inset of Fig.~\ref{fig6}(c). And the position of the second peak gradually shifts to lower fields with increase of temperature. But the second peaks can only exist below 2.2 K. Besides, the second peak effect has not been observed for samples with A15 structure. Therefore, the cause of second peak effect is elusive, which depends on the intrinsic properties of the samples. The MHLs at fixed temperatures are symmetrical around the horizontal axis, indicating a dominant bulk pinning, therefore the Bean critical state model is applicable in this case. From these MHLs, we calculate the critical current density $J_{\rm c}$ based on the formula according to the Bean critical state model $J_{\rm c}$ = $20\Delta M / [a(1-a/3b)] $, as seen in Fig.~\ref{fig6}(c), where $\Delta M$ is the width of MHLs with the unit of emu/cm$^3$, $a$ and $b$ are the width and length of the samples. With increasing temperature, the critical current decreases gradually. At 1.8 K, we have $J_{\rm c} = 1.2 \times 10^4$ A/cm$^2 $ (bcc) and $J_{\rm c} = 5.2 \times 10^4$ A/cm$^2$ (A15) for the bcc and A15 samples, respectively. As we all know, impurities and other defects will form vortex pinning centers, which will hinder the movement of vortices. Therefore, the more flux pinning centers, the greater the critical current of the sample. The large critical current density of the two samples indicates many defects in the samples, which is consistent with the high residual resistivity in our electrical resistive measurement.
	
	\subsection{Scanning tunneling spectrum}	
	\begin{figure*}
		\includegraphics[width=14cm]{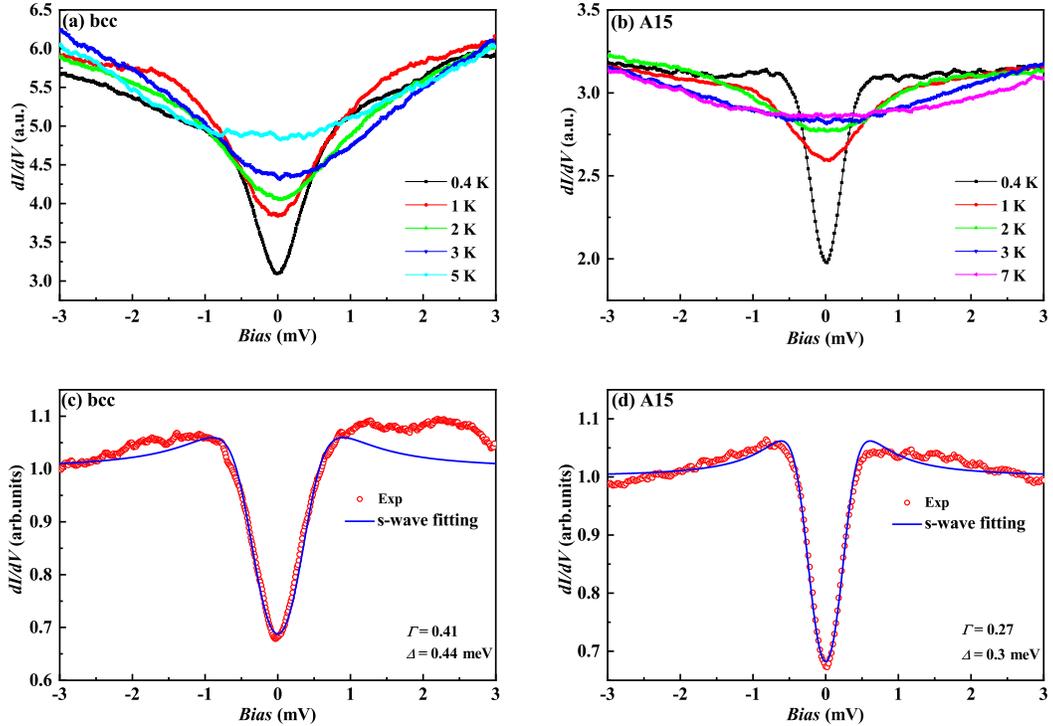}
		\caption{Temperature dependence of the tunneling spectrum (a) on samples with bcc structure and (b) on samples with A15 structure. Tunneling spectrum measured at 0.4 K divided by the normal state background (c) for samples with bcc structure and (d) for samples with A15 structure, respectively (red circles). Each spectrum is fitted by the Dynes model using an s wave superconducting gap function (blue line)} \label{fig7}
	\end{figure*}	

	We further measure the scanning tunneling spectra (STS) on the two samples to investigate the superconducting gap structure. The temperature dependence of the tunneling spectrum for the two samples are plotted in Fig.~\ref{fig7}(a) and~\ref{fig7}(b). It shows that the superconducting gap disappears at around 5 K, close to the critical temperature of the two samples. However, the coherence peaks are blurred and further smeared with the increase of temperature. At $T$ = 0.4 K, the superconducting gap feature with clear suppression of density of states near the Fermi energy can be seen from the spectrum for both samples. Outside the gap, one can see a roughly symmetric background. It can be further illustrated after dividing by the normal state background as can be seen in Fig.~\ref{fig7}(c) and~\ref{fig7}(d). However, one can see that at zero energy, the suppression of the density of states for the two samples is very low. Considering that the samples are not cleaved in situ, we speculate that there could be strong scattering on the sample surface by the impurities or defects. And this could be the reason for the elevated  value of density of states near the Fermi energy. In this case, it seems difficult to determine the pairing symmetry of the superconducting gap for two samples from STS results. However, from the specific heat results, it is konwn that the superconductors could have s wave gaps, thus it is reasonable to use an s wave superconducting gap to fit the data. We thus use the Dynes model to fit our data. The fitting results are also presented in Fig.~\ref{fig7}(c) and~\ref{fig7}(d). It seems that the tunneling spectrum can be fitted quite well for both samples and that we succeed in getting the superconducting gap size, that is 0.44 meV for bcc sample and 0.3 meV for A15 sample. The smeared and suppressed coherence peaks may be induced by the high scattering rate in the samples, since the mean-free pathes are quite small comparting with the coherence length for both samples.

    \subsection{Evolution of superconducting properties at high pressures}

    \begin{figure*}
    \includegraphics[width=15cm]{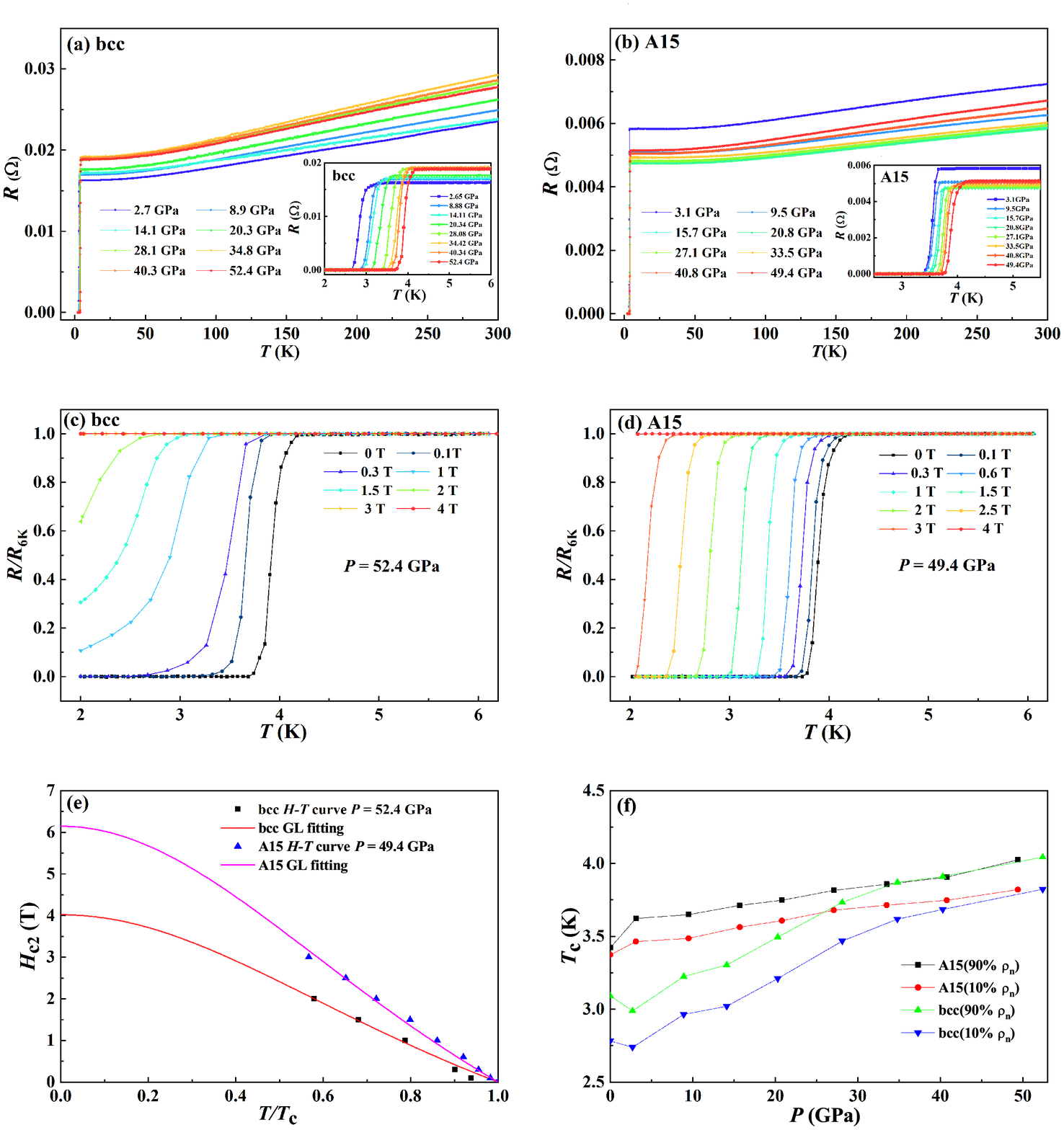}
    \caption{Temperature dependence of resistance for the two kinds of samples at different external pressures (a) for samples with bcc structure and (b) for samples with A15 structure. The insets of (a) and (b) show enlarged views around the superconducting transitions. Temperature dependence of resistance under different magnetic fields (c) at 52.4 GPa for samples with bcc structure and (b) at 49.4 GPa for samples with A15 structure. (e) The upper critical field (symbols) and fitting curves (lines) by the GL theory for the two samples at the highest external pressures. (f) Pressure dependence of superconducting transition temperatures for the two samples.} \label{fig8}
	\end{figure*}	
	Figure~\ref{fig8}(a) and~\ref{fig8}(b) show the temperature dependence of resistance for the two kinds of samples under different external pressures. We can see that the RRR does not change much with increasing pressure, and is always very small, showing a large residual resistance. But the superconducting transition temperature increases gradually with the increase of pressure for the two samples, and the superconducting transition remains steep, as shown in the inset of Fig.~\ref{fig8}(a) and~\ref{fig8}(b). Under the maximum pressure of about 50 GPa, the $T_{\rm c}$ can reach about 4 K. We obtain the values of $T_{\rm c}$ from the $R$-$T$ curves by using criteria of 90$\%$ $R_{\rm n}$ and 10$\%$ $R_{\rm n}$ for all samples and plot them in Fig.~\ref{fig8}(f). Compared with samples with A15 structure, with the increase of pressure, the ratio of increasing $T_{\rm_c}$ is greater for samples with bcc structure. Surprisingly, the relationship of $T_{\rm c}$ and pressure shows a roughly linear behavior and the increase is unsaturated up to the highest pressure in our measurement. In some alloys like NbTi and (ScZrNbTa)$_{0.6}$(RhPd)$_{0.4}$, the $T_{\rm c}$ is robust against large volume shrinkage and shows a monotonous increase with pressure \cite{RSAVSSC,highentropyalloy}. In Fig.~\ref{fig8}(c) and~\ref{fig8}(d), we present the temperature dependence of resistance under different magnetic fields at the highest pressure for both samples. Although the $T_{\rm c}$ of the two samples at the highest pressure are very close, the upper critical fields are different. For samples with bcc structure, with the increase of pressure, the upper critical field of the sample does not increase, which is still about $\mu_{\scriptscriptstyle 0}H_{\rm c \scriptscriptstyle 2}(0)$ = 4.1 T. But for samples with A15 structure, the upper critical field changes from $\mu_{\scriptscriptstyle 0}H_{\rm c \scriptscriptstyle 2}(0)$ = 5.05 T at ambient pressure to $\mu_{\scriptscriptstyle 0}H_{\rm c \scriptscriptstyle 2}(0)$ = 6.15 T at 49.4 GPa.
	
    \section{Discussions}
    For two kinds of Cr$_3$Ru samples with bcc and A15 structures, most of their physical properties show similar unusual behavior. Firstly, very large residual resistivity is found in resistance measurements. The origin of the large residual resistance can be ascribed to the strong scattering in the samples. Considering that the mean free path we obtained for the samples are about 1 $\sim$ 2 nm, which are relatively small compared with the coherence length of about 8 $\sim$ 9 nm. If the grain boundary scattering with the grain size also about 1 $\sim$ 2 nm results in this strong scattering, it is inconceivable that superconductivity is formed in such small grains. Therefore, we tend to think that Cr itself in the system leads to strong scattering, namely, the strong scattering is intrinsic. The slight non-stoichiometry of the Ru and Cr ratios, or the mutual occupation of the Ru and Cr atoms may be the cause for this strong scattering. Secondly, we find strange humps in the $M$(T) curves for both ZFC and FC modes in high temperature region. According to previous reports \cite{CrRu,Cr1-xRux}, the cause of the humps may be the AF ordered phase or the spin fluctuations near phase boundary. Because such humps are less obvious in high field magnetization curves, we tend to attribute it to some AF spin fluctuations which are suppressed under high magnetic field \cite{unconventional}. Thirdly, using the parameters obtained by magnetization and specific heat, the Wilson ratios of the samples are calculated to be larger than that of free-electron ones. Combined with the calculated carrier effective mass, we conclude that there may be moderate electron correlations in these materials. Last but not least, the suppression of the density of states for the samples as observed in tunneling measurements is rather weak at zero energy. This confirms once again there could be strong scattering on the sample surface which diminishes the coherent weight of the Bogoliubov quasiparticles. However, the polycrystalline samples are very pure with only small amount of impurities seen from the XRD patterns. Considering a slight excess of Ru atoms, the original crystal structure may become noncentrosymmetric which brings about such strong scattering. It is widely perceived that unconventional superconductivity may arise in compounds with magnetic elements which yield the magnetic spin fluctuations. We highly suspect that the Cr elements in the present two compounds should also contribute spin fluctuations, as suggested by the non-monotonic temperature dependence of magnetic susceptibility in the normal states. Therefore it is reasonable to believe that the superconductivity found in such a system with strong scattering, moderate electron correlation and AF spin fluctuations is unconventional.

    The two kinds of materials with different structures also show differences in some physical properties. Only the samples with bcc structure exhibit a pronounced second magnetization peak effect below 2.2 K in the MHLs, and the position of the second peak gradually shifts to lower fields with increasing temperature. This difference may be induced by the different microstructures and vortex pinning landscape in these two materials. Besides, only the samples with A15 structure have shown a second specific heat jump in the low temperature region. This anomaly shifts to lower temperatures with increasing magnetic field. We can rule out the possibility that this second specific heat jump is due to some secondary impurity phase, since the XRD patterns show only the tiny impurity phase of Cr oxide which is not superconductive. It is also difficult to attribute this transition to any possible AF order developed below $T_{\rm c}$ since this anomaly can be easily removed by a weak magnetic field. If it would be corresponding to an AF order, a higher magnetic field is needed even the Neel temperature is only 0.85 K. This second specific heat jump may be understood as a phase transition due to unconventional superconductivity. Thus we consider the possibility of a second superconducting transition in this Cr$_3$Ru with A15 structure. According to the theoretical study \cite{secondspecificheatjump}, a hidden transition from an $s + d$ state to an $s + e^{i\eta}d$ state may occur leading to a second jump of specific heat far below $T_{\rm c}$. Due to the magnetic nature of the Cr atoms and the correlation effect, together with the extra amount of Cr beyond the stoichiometry, the system may be non-centrosymmetric, which leads to the existence of, or partially the chiral superconductivity.  This peculiar property may further corroborate that the superconductivity of Cr$_3$Ru is unconventional.

    \section{Conclusions and perspectives}
    In summary, we have successfully synthesized polycrystalline superconducting samples of Cr$_3$Ru with bcc structure and A15 structures. The XRD and EDS analyses confirm the high quality of our samples. And the magnetization and resistivity measurements all exhibit sharp superconducting transitions, with $T_{\rm c}$ = 2.82 K (bcc) and $T_{\rm c}$ = 3.39 K (A15), respectively. The two kinds of samples show many similar unusual physical properties. In magnetization measurements, clear hump structures exist at around 150 K, which are possibly related to antiferromagnetic spin fluctuations. The resistance measurements show a strong residual resistivity with a mean-free-path of about 2 nm for the two samples, indicating an intrinsic strong scattering effect. The low temperature specific heat can be fitted by the s-wave gap for the two samples, the calculated ration $2\Delta/k_{\scriptscriptstyle B}T_{\rm c}$ is around 3.6, indicating a moderate pairing strength. Yet the Wilson ratios are 3.81 and 3.62 for the bcc and A15 samples, suggesting strong electron correlation effect. The two kinds of samples exhibit small lower critical fields and moderate critical current densities. Furthermore, the superconducting transition temperatures both enhance slightly at the external high pressure. Due to strong scattering, the coherence peaks are smeared out on the tunneling spectrum with a gap around 0.3 $\sim$ 0.5 meV. These two kinds of samples with different structures show some difference. A pronounced second peak effect has been observed only in samples with bcc structure below 2.2 K. And only in the samples with A15 structure, we observed the second specific heat jump at about 0.85 K, and it is sensitive to magnetic fields. All these evidences confirm that the Cr-Ru alloy has an unconventional superconductivity, short mean free path and strong scattering. Further studies are highly desired in order to improve the $T_{\rm c}$ higher.

	\section*{ACKNOWLEDGMENTS}
	
We are grateful for the useful discussions with Joerg Schmallian and Ilya Eremin. This work was supported by the National Natural Science Foundation of China (Grant Nos. 11927809 and 12061131001), and the Strategic Priority Research Program of Chinese Academy of Sciences (Grant No. XDB25000000).

\end{document}